\documentclass{desyproc}
\newcommand{\EDW}{\textsc{Edelweiss}}

\begin{document}
\title{Recent status of the Dark Matter search with \EDW}

\author{{\slshape Valentin Kozlov$^1$ on behalf of the EDELWEISS Collaboration}\\[1ex]
$^1$Karlsruhe Institute of Technology, Institut f\"ur Kernphysik, Postfach 3640, 76021 Karlsruhe, Germany}

\contribID{kozlov\_valentin}

\desyproc{DESY-PROC-2012-04}
\acronym{Patras 2012} 
\doi  

\maketitle

\begin{abstract}
The \EDW\ experiment uses Ge-bolometers with an improved background rejection (interleaved electrode design) to search for WIMP dark matter. The setup is located in the underground laboratory, Laboratoire Souterrain de Modane (LSM, France). In 2009-2010 the collaboration successfully operated ten 400-g bolometers together with an active muon veto shielding. Published analysis of this measurement campaign was optimized for WIMP masses above 50~GeV. Recently, the analysis was extended to the low-mass WIMP region using a quality subset of the 2009-2010 data setting new limits on the spin-independent WIMP-nucleon scattering cross-section. We present the low-mass WIMP analysis, background investigations and the latest measurements with a subset of the forty 800-g detectors that will be installed for the {\EDW-III}. Ongoing installation works of the \EDW-III setup and further plans for a next generation experiment, EURECA, are discussed.
\end{abstract}

\section{The \EDW\ experiment and results of the WIMP search for $M_\chi>50$~GeV}

\begin{wrapfigure}{r}{0.52\textwidth}
    \centering
        \includegraphics[width=0.5\textwidth]{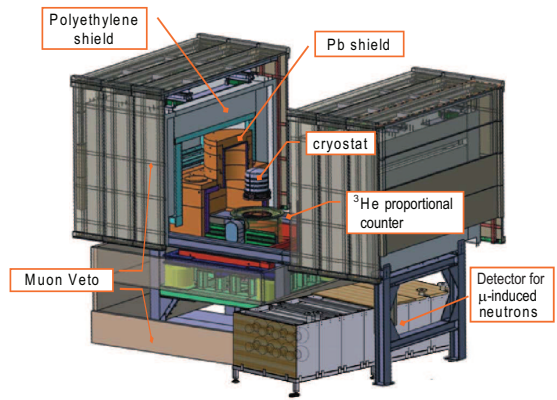}
    \caption{(color online) Layout of the \EDW\ experimental setup with auxiliary detectors: $^3$He proportional counter for thermal neutrons and liquid scintillator detector to measure muon-induced neutrons.}
    \label{fig:edw-setup}
\end{wrapfigure}

The \EDW\ experiment uses Germanium heat-and-ionization bolometers to search for WIMP\footnote{Weakly Interacting Massive Particle} dark matter. It is located in LSM and profits from 4850~m.w.e. shielding, which results in the muon flux of $\sim$5~m$^{-2}$day$^{-1}$. The used Ge-detectors of {\em interdigitized electrode design}~(ID) allow to successfully separate between electron and nuclear recoils as well as between bulk and surface events~\cite{edw09id}. A general overview of the setup is shown in Fig.~\ref{fig:edw-setup}: The central part of the setup is the dilution refrigerator which can host up to 40~kg of detectors at operation temperature of 18~mK. To reduce the external $\gamma$-background a 20~cm lead shield is installed. Neutrons scattered in bolometers produce nuclear recoils, like WIMPs, and thus constitute a prominent background component. A polyethylene layer of 50~cm is used to moderate neutrons while a muon veto system consisting of 100~m$^2$ of plastic scintillators serves to tag muons. Complementary studies of neutron background are performed with a Rn-monitor, $^3$He proportional counters~\cite{edw10he3} and a counter for muon-induced neutrons based on 1~m$^3$ of Gd-loaded liquid scintillator~\cite{edw10nc}. With an ongoing upgrade of the setup (\EDW-III phase) the scientific goal of the \EDW\ experiment is to reach a sensitivity of 5$\cdot$10$^{-9}$~pb for the WIMP-nucleon spin-independent (SI) cross-section within a half year of operation in 2013.


The \EDW\ collaboration successfully operated ten 400-g ID detectors over a period of 14 months from April 2009 to May 2010 and in addition two detectors during an initial run between July and November 2008. The full analysis of the WIMP search for $M_\chi>50$~GeV based on this data set is described in detail in Ref.~\cite{edw11dm}, here we remind main results: A total effective exposure of 384~kg$\cdot$d has been achieved. The average energy resolutions for the detectors were $\sim$1.2~keV (FWHM) for the heat channel and $\sim$0.9 keV~(FWHM) for the ionization channel. The energy threshold for the WIMP search was defined \textit{a priori} as $E_r>20$~keV, above which the efficiency is independent of energy and in order to ensure a maximum exposure in a recoil energy range. Five nuclear recoil candidates were observed above this threshold, while the estimated background is three events. This result was interpreted in terms of limits on the cross-section of SI interactions of WIMPs and nucleons and the value of 4.4$\cdot$10$^{-8}$~pb was excluded at 90\%~C.L. for a WIMP mass of 85~GeV (Fig.~\ref{fig:edw-results}). The \EDW\ and CDMS collaborations also combined their results as the same target material is used in both experiments. This allowed to increase the total data set to 614~kg$\cdot$d equivalent exposure and improve the upper limit on the WIMP-nucleon SI cross-section~\cite{edw11cdms}, e.g. a value of 3.3$\cdot$10$^{-8}$~pb was excluded at 90\%~C.L. for a WIMP mass of 90~GeV where the combined analysis is most sensitive (Fig.~\ref{fig:edw-results}). Chosen methods of combination and further details of the work can be found in Ref.~\cite{edw11cdms}.

\section{Analysis for low-mass WIMPs}
\begin{figure}[!ht]
    \begin{minipage}[c]{0.48\linewidth}
    \centering
    \vspace{8mm}
        \includegraphics[width=0.94\textwidth]{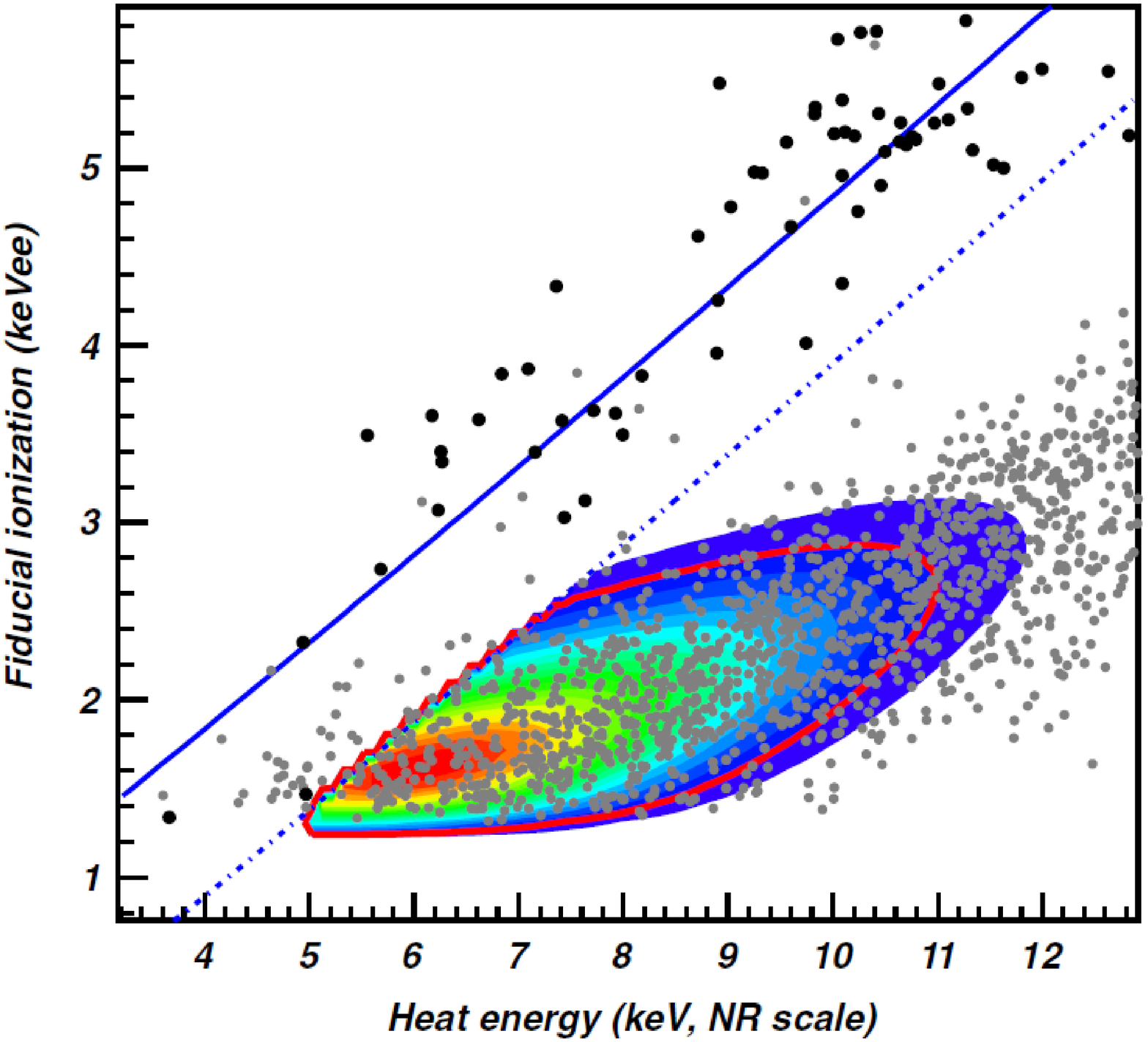} 
    \caption{(color online) Map of the WIMP signal density $\rho(E_r^{(h)}, E_i)$ together with the 90\% WIMP search box (red line) for
a 10~GeV WIMP in one of the detectors. Black dots represent the background, while neutron calibration data is shown with grey dots. The dashed blue line indicates the 95\% $\gamma$-ray rejection~\cite{edw12dm}.}
    \label{fig:signal-density}
    \end{minipage}
    \hspace{2mm}
    \begin{minipage}[c]{0.48\linewidth}
        \centering
        \includegraphics[width=0.98\textwidth]{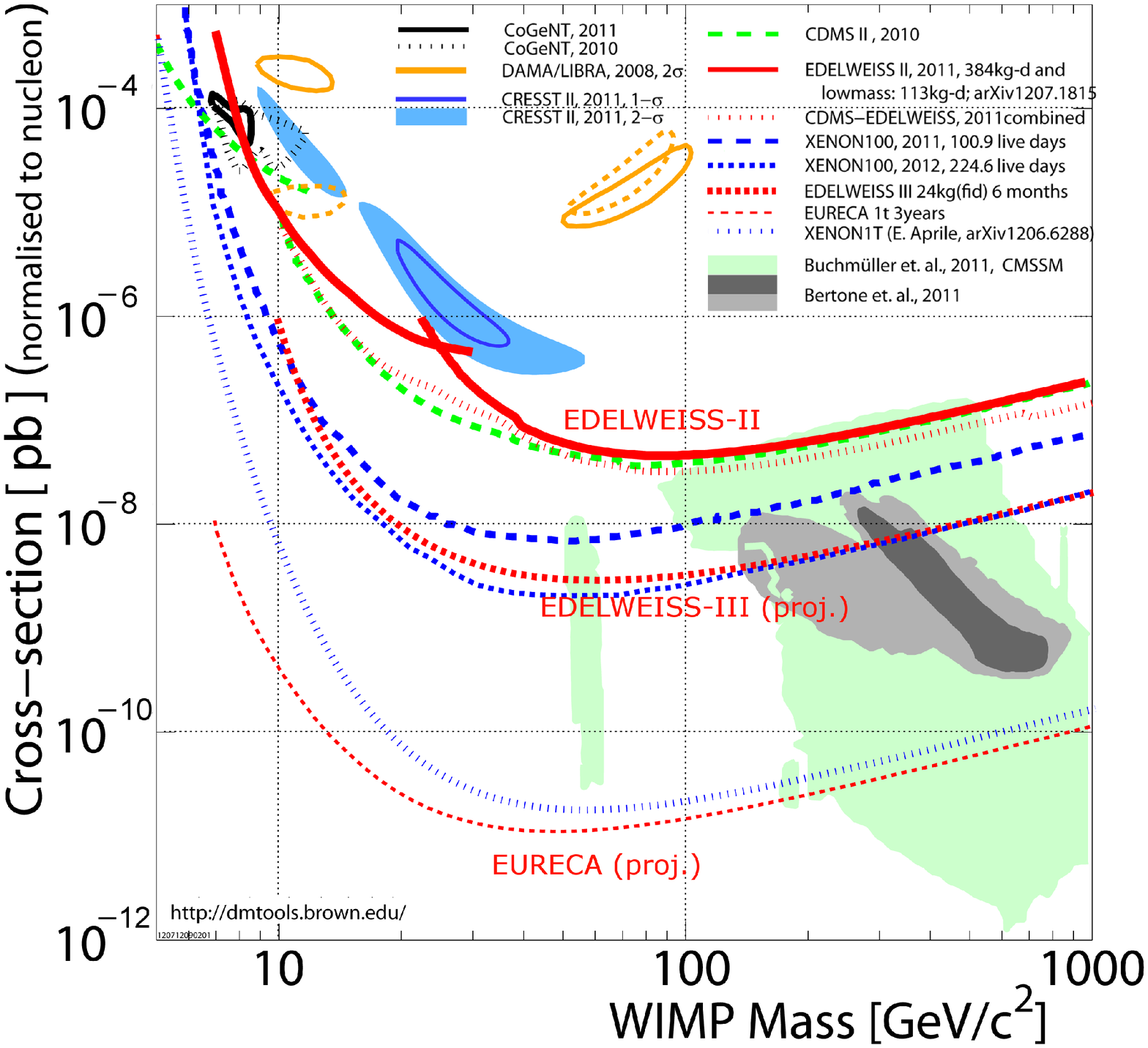} 
    \caption{(color online) The upper limits on the WIMP-nucleon SI cross-section as a function of WIMP mass (the \mbox{\EDW-II} only data are marked with thick solid red lines~\cite{edw11dm,edw12dm}). The green and grey shaded area correspond to theoretical SUSY predictions.}
    \label{fig:edw-results}
    \end{minipage}
\end{figure}
In order to search for WIMPs with a mass of about 10~GeV, one has to go for lower than 20~keV energy threshold as the highest expected recoil energy is of the order of 10~keV. Therefore, one selects a data set on the basis of detector thresholds and backgrounds, for which a low-background sensitivity to nuclear recoils down to 5~keV could be achieved. Here, only a restricted set of 2009-2010 data is used. The low-energy analysis is limited to energies less than 20 keV and thus complementary to and independent from the analyses in Refs.~\cite{edw11dm,edw11cdms}. The data quality cuts result in that only four out of the ten detectors are used for the analysis~\cite{edw12dm}. Among the six rejected detectors, four had nonfunctioning electrodes and/or poor resolutions for one or several channels, one was rejected due to a relatively intense accidental source of $^{210}$Pb in its vicinity, and one had a low-energy $\gamma$-ray background four times larger than that observed on other detectors. The  event-based quality cuts are applied for the event reconstruction and identical to that described in Ref.~\cite{edw11dm}. As well the events in coincidence with any other bolometer in operation or muon veto are rejected. However, the estimate of the recoil energy, $E_r^{(h)}$, is done differently from Refs.~\cite{edw11dm,edw11cdms}, i.e. it is determined solely from the heat channel signal and corrected for the Neganov-Luke effect assuming that events are due to nuclear recoils. The advantage of such estimator is a better energy resolution on $E_r^{(h)}$ for nuclear recoils than the estimator used in Refs.~\cite{edw11dm,edw11cdms}. The efficiency of the online DAQ trigger over time as a function of $E_r^{(h)}$ is estimated from the resolution of the heat channel and the adaptive trigger threshold, and validated using the flat, low-energy Compton plateau in $\gamma$-ray calibration data. The rejection of nonfiducial interactions and ionizationless events is based on signals from the ionization channels. The efficiencies of the offline analysis cuts and the measured resolutions of fiducial ionization energy, $E_i$, and heat energy, $E_r^{(h)}$, are then used to construct for a given detector and a given WIMP mass a WIMP signal density in the $(E_r^{(h)}, E_i)$ plane (Fig.~\ref{fig:signal-density}). A WIMP search region in this plane is defined as the region containing 90\% of all the calculated WIMP signal density below the $\gamma$-rejection cut. The sensitivity to the WIMP scattering signal is demonstrated by neutron calibration data. For each WIMP mass, the 90\%~C.L. upper-limit cross-section is calculated by integrating the WIMP signal density $\rho(E_r^{(h)}, E_i)$ over the WIMP search region and assuming Poisson statistics. The obtained limit significantly improves the \EDW\ sensitivity below $M_\chi=20$~GeV and excludes the regions favored by DAMA/LIBRA, CRESST and part of the CoGeNT space above $M_\chi=8$~GeV (Fig.~\ref{fig:edw-results})~\cite{edw12dm}.

\newpage
\section{Outlook: Improvements towards better sensitivity}

\begin{wrapfigure}{r}{0.42\textwidth}
   \centering
        \includegraphics[width=0.4\textwidth]{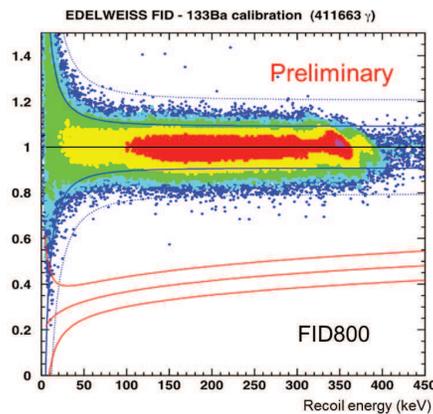}
        \caption{(color online) Ionization yield (ratio between the ionization energy and the recoil energy) versus energy after fiducial cuts, obtained with new FID800 detectors using the $^{133}$Ba $\gamma$-source.}
    \label{fig:qplot-fid800}
\end{wrapfigure}

Further improvements of the setup concern nearly all aspects of the experiment:  New detectors with interleaved electrodes covering also the lateral surfaces of the crystal have been developed, the so-called  {\em fully interdigitized} (FID) bolometers with twice the mass (800~g) and increased fiducial volume ($\sim$600~g). These detectors confirmed to have better rejection of $\gamma$-events (Fig.~\ref{fig:qplot-fid800}) and forty of them are planned for installation in the upgraded \EDW\ setup. The energy resolution of the ionization channels should also improve due to a redesign of the front-end electronics and identification of external noise sources. An additional polyethylene shield will be placed between the lead layer and the cryostat, which is expected to reduce the flux of ambient neutrons by an order of magnitude. New cabling and materials with lower level of radioactivity should reduce the $\gamma$-background up to a factor of 6. Better characterization of the muon veto together with supplementary muon veto modules suppose to improve rejection of muon induced neutrons by an order of magnitude. Modifications of DAQ go towards better scalability and integration of various readouts. Also new analysis tools~\cite{edw12kdata} are being developed to facilitate the monitoring and analysis of significantly enlarged data sets. The goal of the funded \EDW-III project is to acquire an exposure of 3000~kg$\cdot$d within a half year of operation in 2013 to reach a WIMP-nucleon scattering cross-section sensitivity of ~5$\cdot$10$^{-9}$~pb or better. The ongoing research together with detector R\&D and the detailed studies of the background conditions in LSM put a good base for a dark matter experiment of next generation, EURECA~\cite{eureca07}, a 1-ton cryogenic detector array to probe WIMP-nucleon SI interaction down to $10^{-11}$~pb (Fig.~\ref{fig:edw-results}).

\paragraph{Acknowledgments:} This work is supported in part by the German ministry of science and education (BMBF), by the Helmholtz Alliance for Astroparticle Physics, by the French Agence Nationale pour la Recherche, by Science and Technology Facilities Council (UK) and the Russian Foundation for Basic Research(grant No. 07-02-00355-a).
 

\begin{footnotesize}

\end{footnotesize}


\end{document}